\documentclass[british]{IOS-Book-Article}     
\usepackage{float}
\usepackage{mathptmx}
\usepackage{amsmath}
\usepackage{graphicx}
\usepackage{algpseudocode}
\usepackage{algorithm}

\begin{document}
\begin{frontmatter}          
%
\title{An Efficient Thread Mapping Strategy\\ for Multiprogramming on\\ Manycore Processors}
\runningtitle{}

%
\author{\fnms{Ashkan} \snm{Tousimojarad}},
\author{\fnms{Wim} \snm{Vanderbauwhede}}
\runningauthor{}
\address{School of Computing Science\\ University of Glasgow}
\begin{abstract}
The emergence of multicore and manycore processors is set to change
the parallel computing world. Applications are shifting towards
increased parallelism in order to utilise these architectures efficiently.
This leads to a situation where every application creates its desirable
number of threads, based on its parallel nature and the system resources
allowance. Task scheduling in such a multithreaded multiprogramming
environment is a significant challenge. In task scheduling, not only the order of the execution, but also the mapping
of threads to the execution resources is of a great importance. In
this paper we state and discuss some fundamental rules based on results obtained
from selected applications of the BOTS benchmarks on the 64-core TILEPro64
processor. We demonstrate how previously efficient mapping policies
such as those of the SMP Linux scheduler become inefficient when the
number of threads and cores grows. We propose a novel, low-overhead
technique, a heuristic based on the amount of time spent
by each CPU doing some useful work, to fairly distribute the workloads
amongst the cores in a multiprogramming environment. Our novel approach
could be implemented as a pragma similar to those in the new task-based
OpenMP versions, or can be incorporated as a distributed thread mapping
mechanism in future manycore programming frameworks. We show that
our thread mapping scheme can outperform the native GNU/Linux thread
scheduler in both single-programming and multiprogramming environments.
\end{abstract}


\end{frontmatter}


\section*{Introduction}

Recently, task parallelism has gained a great importance in parallel
computing. Some task parallel frameworks provide explicit task parallelism
\cite{ayguade2009design}, while in others parallelism is expressed
implicitly \cite{tousimojarad2013parallel}. The OpenMP \textit{task}
directives can be used to define units of independent work as \textit{task}s.
However, the scheduling decisions are leaved to the runtime systems.
\cite{olivier2010comparison} suggests the improvement of OpenMP runtime
support. Nanos v4 \cite{teruel2007support} is an OpenMP runtime library
that provides some mechanisms to allow the user to choose between
different task scheduling policies.  In \cite{duran2008evaluation},
two families of the schedulers are added to the Nanos runtime: Breadth-First schedulers
and Work-First schedulers. The main focus is on how threads execute
the tasks, while our focus in this paper is on where to place those threads.

Task scheduling is the arrangement of tasks in time and space on the
available execution resources. Therefore, not only the order of execution,
but also optimum mapping of all threads to the existing resources
is crucial to get the best result. In \cite{zhang2008extending},
binding threads to the physical processors, also referred to as \textit{thread
affinity}, is proposed to be a part of the OpenMP standard. It has
been shown how a simple Round-Robin mapping scheme can improve the
performance.

An optimal solution to task scheduling problem cannot be found in
polynomial time which means it is an NP-hard problem. This motivates
researchers to develop heuristics to find near optimal solutions \cite{sinnen2007task}.
Previous studies show that efficient mapping of threads to specific
cores, providing load balancing among the cores would result in improved
performance. In \cite{sasaki2012scalability}, a sophisticated scheduling
scheme in multiprogramming environment is developed, based on dynamically
prediction of the applications' scalability. It motivated us by highlighting
this fundamental problem that the performance of some applications
without linear scalability tend to decrease when more number of cores
are allocated to them. Limiting the number of threads is a way to
overcome this issue. Therefore, for an application with smaller number
of threads than the number of cores, it becomes important to determine
where to map those threads. We will address this issue by calculating
the current CPU load of the tiles before mapping the threads to them.

Generally, scheduling policies can be evaluated based on system-oriented
or user-oriented criteria \cite{stallings2012operating}. A system-oriented
metric is based on the system's perspective and quantifies how effectively
and efficiently the system utilises the resources, while the focus
of a user-oriented metric is on the behaviour of the system from user's
perceptive, e.g. how fast a single program is executed. An example
of system-oriented metrics is \textit{throughput}, which is the number
of programs completed per unit of time. \textit{Turnaround time} is
an example of user-oriented metrics, which is the time between submitting
a job and its completion. In this work, we have used \textit{Turnaround
time} to evaluate different mapping techniques on the TILEPro64 machine.

\section{Selected Benchmarks}
We show how different thread mapping strategies can affect the
performance of four benchmarks from the Barcelona OpenMP Tasks Suite (BOTS) \cite{duran2009barcelona},
selected for their different characteristics. The mapping techniques are low-overhead,
and can be combined with different cut-off strategies and applied
on either \textit{tied} or \textit{untied} tasks.  

It is important to note that the applications which do not scale very
well are more challenging for parallel computing. Embarrassingly parallel
algorithms are easy to parallelise since the tasks are completely
(or almost) independent. They can easily run on different processing
cores without the need to share data or exchange any information with
each other. We have used a benchmark that scales approximately linearly
(NQueens), one that does not scale well when the number of threads
grows (Strassen), and two others that reach their saturation phases
(Sort and Health). The input sets are chosen in such a way that the turnaround times of the programs range from a few seconds to a few tens of seconds. The aim is to show that the overhead of the proposed mapping technique
is negligible, even for programs with small turnaround times. 

The target platform is the TILEPro64, which runs Tile Linux that is based on the standard open-source Linux
version 2.6.26. The C compiler used is the one provided in the Multicore Development Environment (MDE) 3.0
from Tilera Corporation, which is called Tile-cc and is based on the GCC 4.4.3. The only change made to the BOTS 1.1.2 configuration file is the name of the compiler. 

\begin{enumerate}
\item \textbf{Sort (untied):} Sorts a random permutation of \textit{n} 32-bit numbers with a fast parallel
sorting variation of the ordinary merge sort. First, it divides an
array of elements in two halves, sorting each half recursively, and
then merging the sorted halves with a parallel divide-and-conquer
method rather than the conventional serial merge. Tasks are used for
each split and merge. When the array is too small, a serial quick
sort is used to increase the task granularity. We have used the default
cut-off values (2048) when sorting an array of 50M integers. To avoid
the overhead of quick sort, an insertion sort is used for small arrays
of 20 elements.
\item \textbf{Health (manual-tied):} This program simulates de Columbian Health Care System. Each element
in its multilevel lists represents a village with a list of potential
patients and one hospital. The status of a patient in the hospital
could be waiting, in assessment, in treatment, or waiting for reallocation.
Each village is assigned to one task. The probabilities of getting
sick, needing a convalescence treatment, or being reallocated to an
upper level hospital are considered for the patients. At each time-step,
all patients are simulated according to these probabilities. To avoid
indeterminism in different levels of the simulation, one seed is used
for each village. Therefore, all probabilities computed by a single
task are identical across different executions and are independent
of all other tasks. three different input sizes are available in the benchmark
suite. We have used them in different scenarios. However, the performance
scalability of the single program is presented using the medium-size
input.
\item \textbf{Strassen (tied):} The Strassen algorithm employs a hierarchical decomposition of a matrix
for multiplication of large dense matrices. Decomposition is performed
by dividing each dimension of the matrix into two parts of equal size.
For each decomposition a task is created. A matrix size of $2048\times 2048$
is used for the purposes of this experiment.
\item \textbf{NQueens (manual-untied):} The NQueens benchmark computes all solutions of the n-queens problem,
whose aim is to find a placement for \textit{n} queens on an $n \times n$ chessboard such that none of the queens attack any other.
It uses a backtracking search algorithm with pruning. A task is created
for each step of the solution,and it has an almost linear speed-up.
\end{enumerate}

\section{Mapping Strategies}
We have performed the experiments with four different mapping strategies.
In the first part of the experiments, we measure the execution time
of each benchmark under these mapping policies in a single-programming
environment. The second part is to investigate the behaviour of these mapping strategies in a multiprogramming environment. In each of our mapping schemes,
every thread decides about its mapping itself. It first finds a suitable
core, maps itself to it and starts doing some work or goes to sleep.
In the OpenMP code, this happens after the \textit{parallel} keyword,
which is the point where the thread creation happens. 

\subsection{Linux Scheduler} 
The first option is to leave any scheduling
decision to the native Linux scheduler. Tilera's version of SMP
Linux, called Tile Linux is based on the standard open-source Linux
version 2.6.26. The default scheduling strategy in Linux is a priority-based
dynamic scheduling that allows for thread migration to idle cores
in order to balance the runqueues. 

Having a single runqueue for all processors in a Symmetric Multiprocessing
(SMP) system, and using a single runqueue lock were some of the drawbacks
of the Linux 2.4 scheduler. Linux 2.6 implemented a priority-based
scheduler known as the O(1) scheduler, which means the time needed
to select the appropriate process and map it to a processor is constant.
One runqueue data structure per each processor keeps track of all
runnable tasks assigned to that processor.

At each processor, the scheduler picks a task from the highest priority
queue. If there are multiple tasks in that queue, they are scheduled
in a Round-Robin manner. There is also a mechanism to move tasks from
the queues of one processor to those of another. This is done periodically
by checking whether the  \textit{cpu\_load} is imbalanced.
In the Linux terminology, \textit{cpu\_load} is the average of the
current load and the old load. The current load is the number of active
tasks in the CPU's runqueue multiplied by \texttt{SCHED\_LOAD\_SCALE},
which is used to increase the resolution of the load \cite{aas2005understanding}.
What we will refer to as \textit{load} in our proposed technique is
the amount of time spent in each processor doing some useful work.

\subsection{Static Mapping} 
In the static mapping strategy, threads are pinned
to the processing cores based on their \textit{thread\_id}s in an
ordered fashion. The decision is taken at compile time, which would
cause an obvious disadvantage: It cannot tune itself with multiprogramming,
since every program follows the same rule, and if the number of threads
are less than the number of cores, then some cores get no threads
at all. It might be discussed why at the first place, the number of
threads in each program should be less than the number of cores. The
answer to this question can be found in the applications which do
not have linear speed-up, and after a certain number of threads reach
their saturation phase. An example is the Sort program in this benchmark
suite. This behaviour is reported in \cite{sasaki2012scalability} as
well, where the authors emphasise that the scalability of some programs
tend to saturate at some points, and their performance is degraded
by adding more cores. 
 
\subsection{Basic Lowest Load (BLL)}
The Lowest Load mapping technique is presented as two different methods. 
The first one, assumes the term
\textit{load} as an equivalent to a thread. Therefore, if one thread
is mapped to a core, the core's load becomes 1. We call this method
\textit{Basic Lowest Load} (BLL). It fills out the cores of the system
in a Round-Robin fashion. Again, this technique is not aware of what
is going on in the system. There are many situations in which some
idle cores are ignored. We will show an example in the next section. 

\subsection{The Extended Lowest Load (XLL)}
The XLL gets the cores' information from the \textit{/proc/stat} file
in Linux. The amount of time each core has done different types of work is specified
with a number of time units. The time units are expressed in USER\_HZ or Jiffies, which are typically hundredths of a second. The number of Jiffies in user mode is selected as \textit{load}. In this
technique, every thread scans the current \textit{load}s of the cores. It then searches for a core with the least change
from its old \textit{load }value. The thread maps itself to that core
and starts working. In other words, the actual target of this policy is the least busy core.
Except from its dynamic awareness of the system,
another difference with the BLL becomes highlighted when a thread
is created but goes instantly to the sleeping mode. The XLL automatically
finds the sleeping threads since they do not produce any \textit{load},
and hence more threads can be assigned to the corresponding cores,
while the BLL only counts the number of pinned threads to the core,
no matter if they are sleeping or doing some work. 
The algorithm for the XLL methodology is as follows:

\begin{algorithm}
\caption{The XLL Methodology}
\label{array-sum}
\begin{algorithmic}[1]
\Procedure{FindBestTarget}{}
  \State GetTheLock();
  \For {each int i in Cores }
	\State Scan(CurrentLoad[i]); \textit{\Comment{Scans from the /proc/stat file}}
	\State Cores[i].change = CurrentLoad[i] - Cores[i].load + Cores[i].pinned;
	\State Cores[i].load = CurrentLoad[i] + 10; \textit{\Comment{Creates a better resolution}}
 \EndFor
   \For {each int i in Cores }
	\If {Cores[i].change  \textless  Cores[BestTarget].change} 
	\State BestTarget = i; \textit{\Comment{Finds the least busy core}}
	\EndIf
  \EndFor
  \State SetAffinity(BestTarget);
  \State Cores[BestTarget].pinned++; \textit{\Comment{Increments the number of pinned threads}}
  \State ReleaseTheLock();
\EndProcedure
\end{algorithmic}
\end{algorithm}

The proposed methodology requires a globally shared data structure that keeps track of the system's cores. This data structure can be implemented in a runtime system as in our work, or can be embedded in the Linux kernel. It is worth mentioning that this methodology is portable across similar multicore/manycore platforms.

\section{Results}
\begin{figure}[h]
\begin{centering}
{\footnotesize }%
\begin{tabular}{cc}
\includegraphics[clip,width=0.45\textwidth]{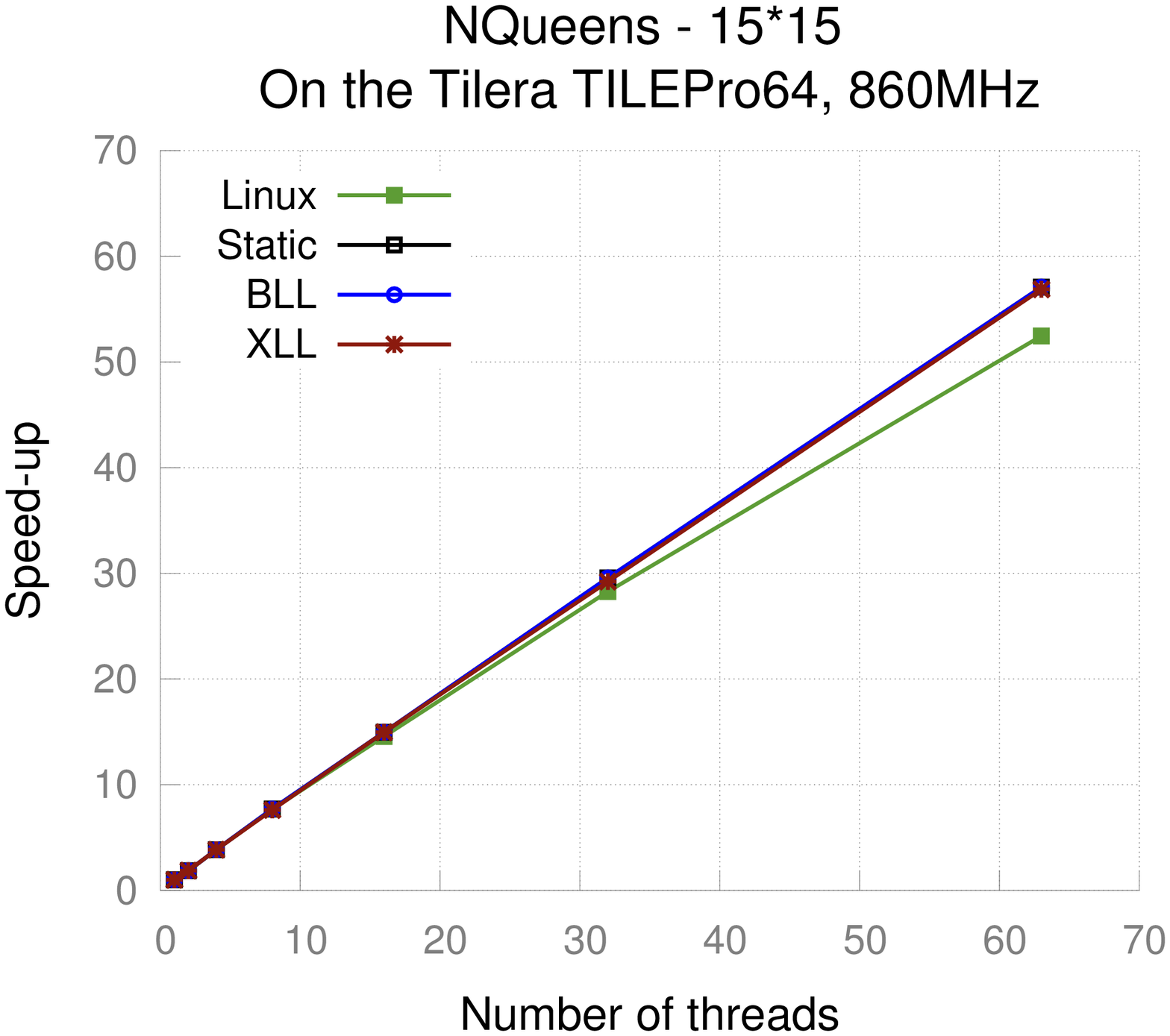} & \includegraphics[clip,width=0.45\textwidth]{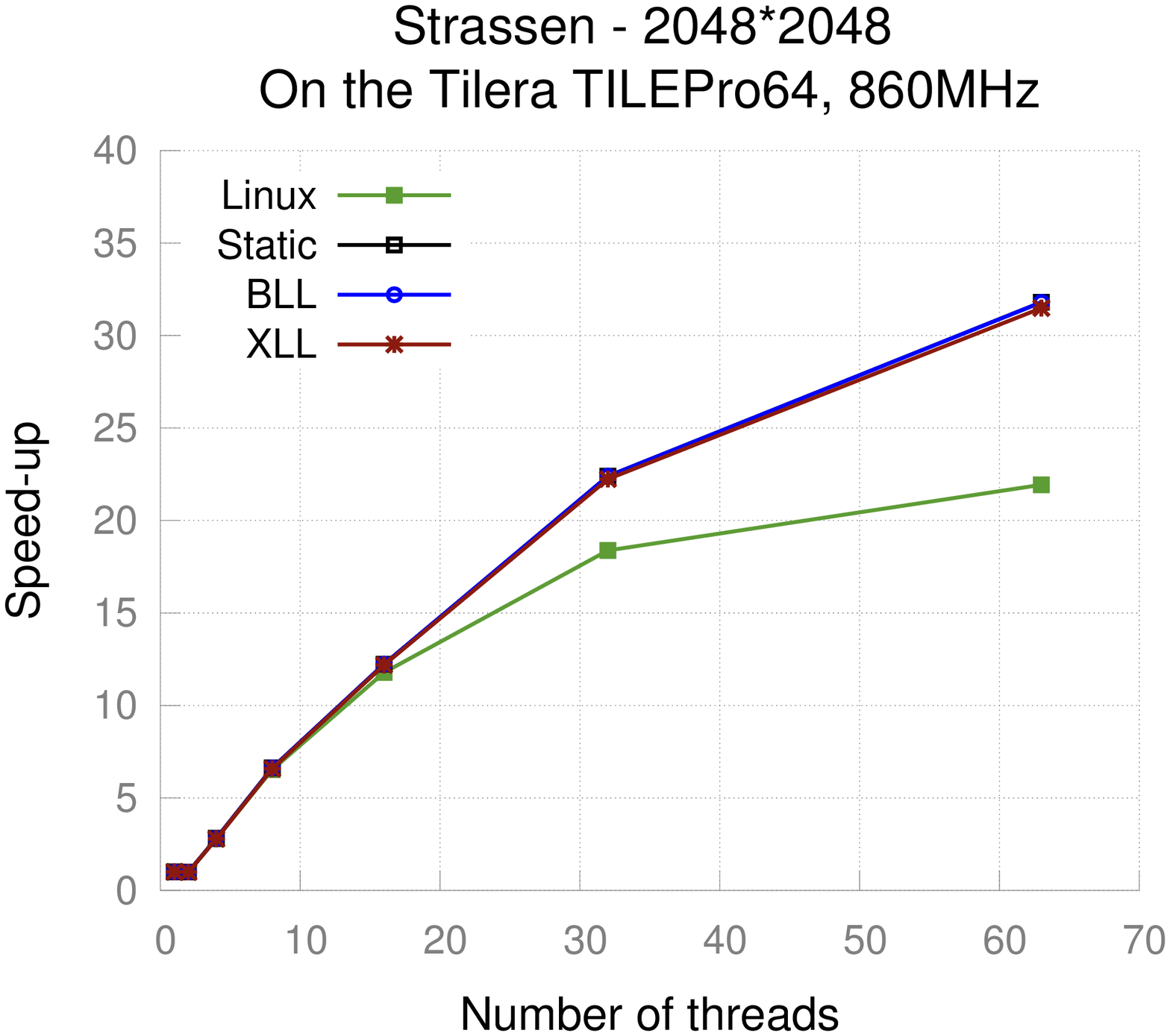}\tabularnewline
{\footnotesize (a)} & {\footnotesize (b)}\tabularnewline
\end{tabular}
\par\end{centering}{\footnotesize \par}

\begin{centering}
{\footnotesize }%
\begin{tabular}{cc}
\includegraphics[clip,width=0.45\textwidth]{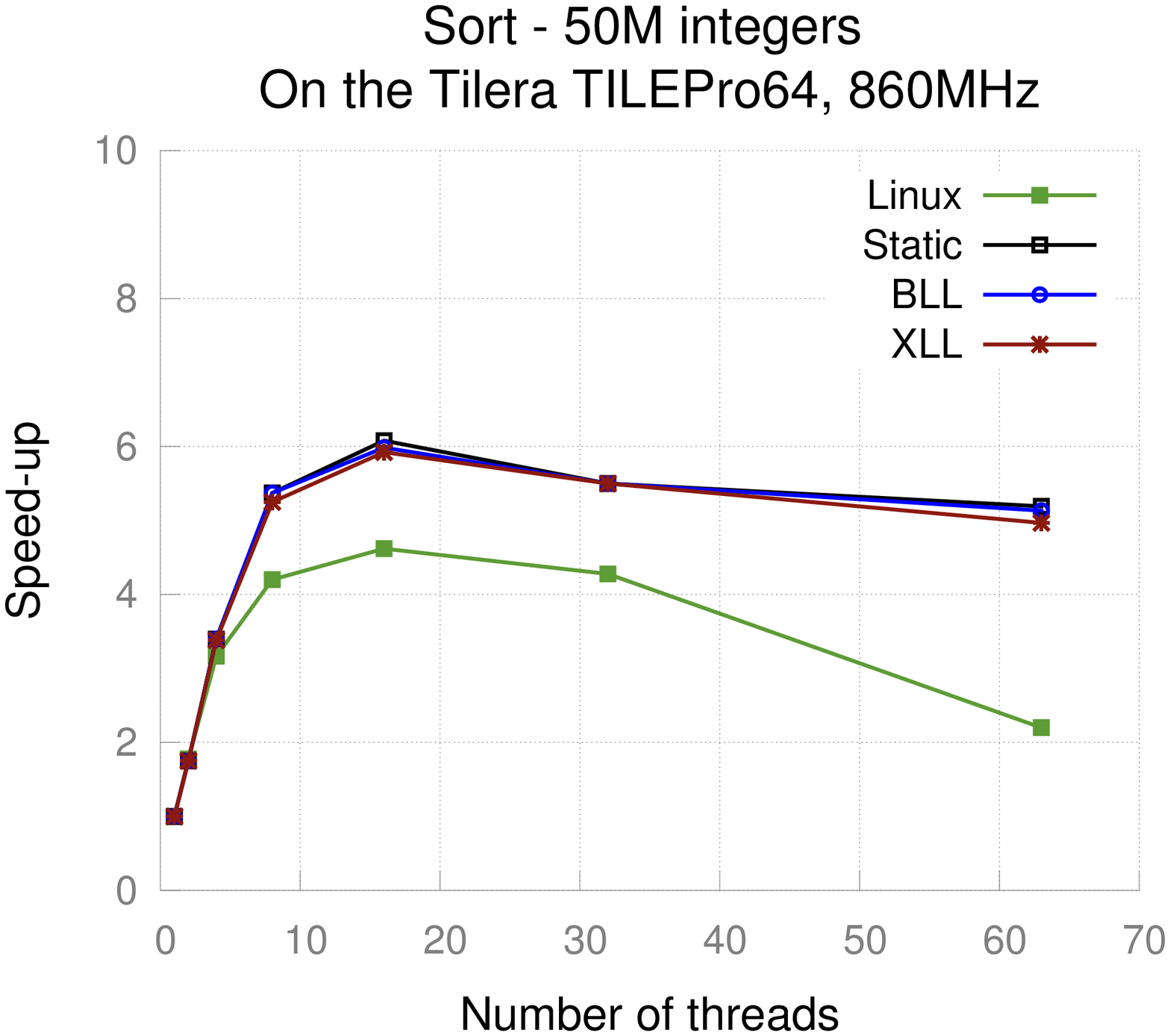} & \includegraphics[clip,width=0.45\textwidth]{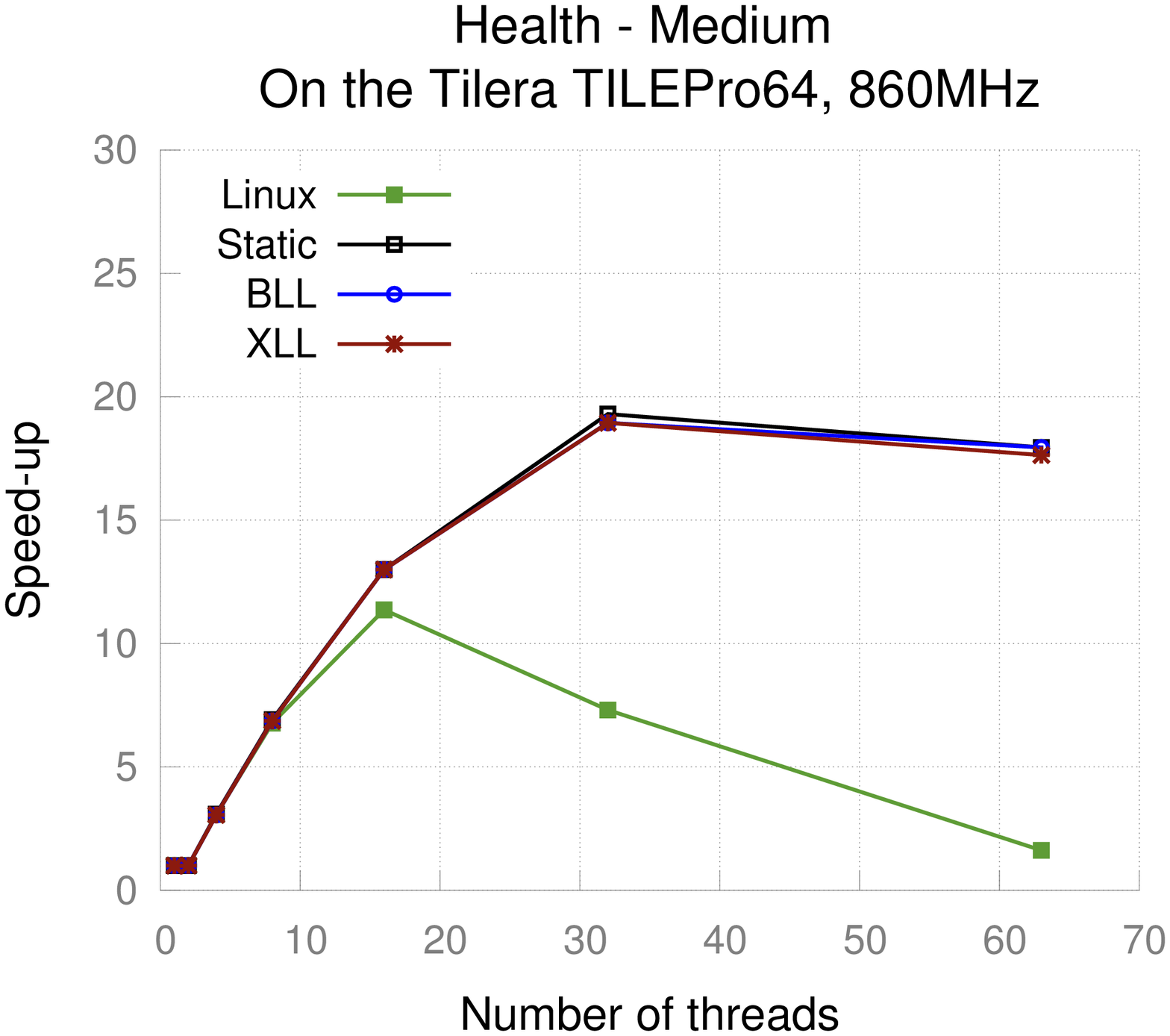}\tabularnewline
{\footnotesize (c)} & {\footnotesize (d)}\tabularnewline
\end{tabular}
\par\end{centering}{\footnotesize \par}

\caption{\label{fig:4-bots-benchmarks}Speed-up of the selected benchmarks under different
mapping policies (a) NQueens $15\times 15$ board, (b) Strassen $2048\times 2048$
matrix, (c) Sort 50M integers, (d) Health Medium input}
\end{figure}

The first step is to analyse the performance behaviour of each benchmark individually.
 Figure \ref{fig:4-bots-benchmarks} shows how
our 4 applications from the BOTS benchmark suite scale on the TILEPro64.
As mentioned before, not all the applications benefit from the increased
number of threads. Thus, in order to get the best performance, one
solution is to limit the number of threads based on the scalability
of the application. 

For a single-program workload, The Static, BLL, and XLL techniques
behave very similarly. The comparison with the Static mapping verifies
that the overhead of the BLL and XLL techniques are negligible. Moreover,
we can see that even for a single-program workload, the proposed mapping
technique (XLL) works better than Linux. 

\begin{figure}[h]
\begin{centering}
\includegraphics[width=0.60\textwidth]{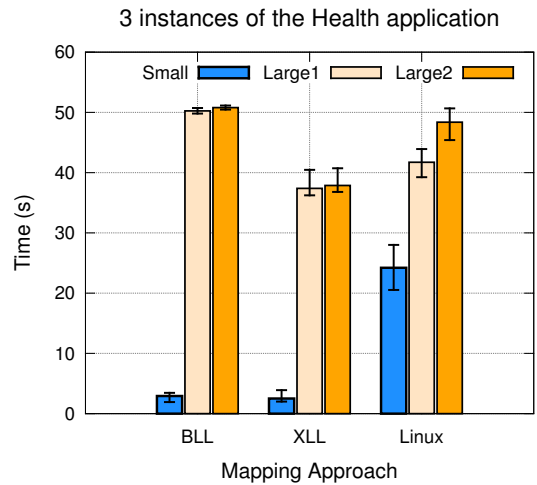}
\par\end{centering}

\caption{\label{fig:3-3-health}The first scenario: The inefficiency of the
BLL}
\end{figure}

For multiprogram workloads, we have considered three different scenarios
to show how the XLL mapper can result in better performance. It is
worth mentioning that each experiment was run 20 times.

First, we have to show why BLL, which is a simple Round-Robin mapping
algorithm is inefficient. For this purpose, we have considered three Health
programs, each of which with 32 threads (that gave us the best performance
in the single-program scenario). Two programs have large inputs and
one has a small input. The programs enter the system with the interval
of 6 Secs. We have previously discussed why Static mapper cannot handle
multiprogramming scenarios. The inefficiency of the BLL is also evident
from Figure \ref{fig:3-3-health}.

The first scenario clearly shows that the XLL is the winning policy.
The scenario is designed in such a way that the program with the small
input data set finishes before the second large program enters the
system. In the case of BLL, the threads of the first large program
are mapped to the first 32 cores of the system. The threads of the
small program are mapped to the last 31 cores plus the first core
(there are 63 cores to use). Then the small program finishes and the
second large program enters the system, but the BLL cannot use the
recently freed cores. Instead, based on the Round-Robin algorithm,
the threads of the second large program are mapped to the cores 2
to 33, while most of these cores (except one of them) are already busy serving
the first large program.

\begin{figure}[h]
\begin{centering}
\includegraphics[width=0.60\textwidth]{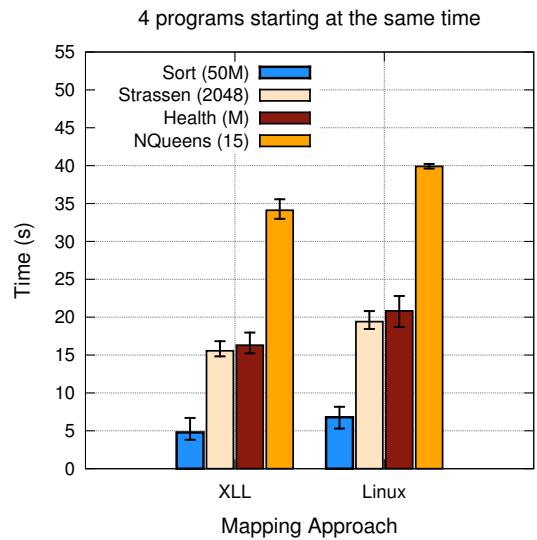}
\par\end{centering}

\caption{\label{fig:4-same-time}The second scenario: Running selected programs
as the same time}
\end{figure}

The second scenario is to run all four programs selected from the BOTS
at the same time. Although they start at the same
time, their thread creation time is different. This is due to the
fact their initialisation phases and memory allocation times are different.
Recall that thread creation happens whenever the execution reaches
the \textit{parallel} keyword in the OpenMP code. According to the
single-program performance, the Sort program (50M integers) is limited
to 16 threads, the Health program (Medium input) is limited to 32
threads, and both Strassen ($2048\times 2048$) and NQueens ($15\times 15$) are
run with 63 threads. The result is shown in Figure \ref{fig:4-same-time}.

Once again, the XLL results in better performance. The turnaround
times for all 4 programs are smaller when the XLL policy is applied.

\begin{figure}[h]
\begin{centering}
\includegraphics[width=0.60\textwidth]{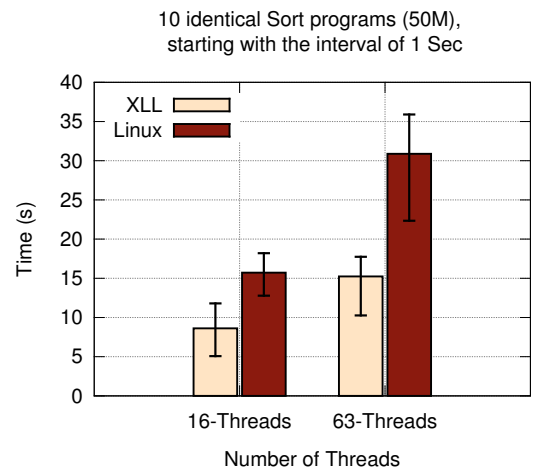}
\par\end{centering}

\caption{\label{fig:10-sort}The third scenario: Running 10 identical instances
of the Sort program }
\end{figure}

The third scenario gives a better insight on how the XLL mapper outperforms
the Linux scheduler when the system is busy. For this scenario, we
have used 10 identical instances of the Sort program arriving the
system one after the other with the interval of 1 second. The result
is depicted in Figure \ref{fig:10-sort}.

Figure \ref{fig:10-sort} shows that the results with both policies
are significantly better when each Sort program uses 16 threads rather
than 63. It again verifies that increasing the number of threads does
not necessarily result in better performance. It is also evident how
much our novel XLL mapping technique can outperform the native Linux
scheduler in a multiprogramming environment.

\section{Conclusion}
In this work, we have performed an analysis of multiprogramming performance
on a Tilera manycore system using the BOTS benchmark. We have shown
that although the current SMP Linux scheduler performs well for small
numbers of threads (up to 8 threads), the scheduling performance
degrades for increasing numbers of cores and threads, leading to poor performance
on manycore systems. We observe that increasing the number of threads does not necessarily lead to better performance
and can even degrade the overall performance. With a smaller
number of threads than the number cores, different mapping configurations
are possible, which allows to optimise the performance. 

We have presented a novel, low-overhead, fundamental mapping strategy
to provide load balancing in a multithreaded multiprogramming environment:
the \textit{Extended Lowest Load} (XLL) technique uses a heuristic
to find the optimal target core for each thread. Although this work
is in an early stage, the results are very promising. In this paper,
we have shown how our XLL mapping scheme outperforms the native GNU/Linux
thread scheduler in different scenarios.


\bibliographystyle{unsrt}
\bibliography{Parco}

\end{document}